# Variable p norm constrained LMS algorithm based on gradient of root relative deviation

Yong Feng✉, Fei Chen and Jiasong Wu

A new $L_p$-norm constraint least mean square ($L_p$-LMS) algorithm with new strategy of varying $p$ is presented, which is applied to system identification in this letter. The parameter $p$ is iteratively adjusted by the gradient method applied to the root relative deviation of the estimated weight vector. Numerical simulations show that this new algorithm achieves lower steady-state error as well as equally fast convergence compared with the traditional $L_p$-LMS and LMS algorithms in the application setting of sparse system identification in the presence of noise.

*Introduction:* Recent years have brought up an avalanche of interest in sparse adaptive filtering in a variety of applications, such as sparse channel estimation and system identification, and such a tilde is mainly stimulated by the study of the least absolutely shrinkage and selection operator (LASSO) and compressed sensing (CS). Among these methods, the least mean square (LMS) algorithm is probably the most notable and extensively used on the account of its computational simplicity and ease of implementation. Moreover, the family of norm constrained LMS algorithms have attracted much attention and are able to gain better performances than the traditional LMS approach and greater robustness against additive noise in identifying a sparse system [1-4].

So many norm constrained LMS algorithms have been released to better the performance of the classical LMS algorithm, such as the $L_0$-norm penalty LMS ($L_0$-LMS) [1], $L_1$-norm constrained LMS ($L_1$-LMS) [2], $L_p$ norm LMS ($L_p$-LMS) [3] and $L_p$ norm-like LMS ($L_{pl}$-LMS) [4]. Different corresponding norm constraints are integrated to the loss function of the classical LMS in the aforementioned new methods in order to accelerate the convergence and / or lower the steady state mean square error (MSE). However, most of these existing algorithms as well as their variants have the difficulty of adaptability to the sparsity of the unknown systems to be estimated, due to the lack of any adjustable factors [4].

In this letter, we derive a new $L_p$-LMS algorithm with a varying $p$ to address the aforementioned concern, which is achieved by iteratively adjusting $p$ with a gradient method applied to the root relative deviation of the estimated weight vector. Hence it results in a better $L_p$-norm penalty for the proposed $L_p$-LMS algorithm. Moreover, the numerical simulation results show that the proposed algorithm has lower MSE than the classical LMS and $L_p$-LMS algorithms.

*Conventional $L_p$-LMS Algorithm:* Let $y_k$ stand for the output of a finite impulse response system which has an additive noise $n_k$ at time $k$, represented as

$$y_k = \mathbf{w}^T \mathbf{x}_k + n_k \tag{1}$$

where $\mathbf{w}$ is the weight vector of length $N$ and the impulse response of an unknown system, and $(\cdot)^T$ denotes the transpose operator. $\mathbf{x}_k$ represents the stationary input vector with zero mean and autocorrelation matrix $\mathbf{R}$, which consists of the last $N$ input signal samples, i.e., $\mathbf{x}_k \triangleq [x_k, x_{k-1}, \cdots, x_{k-N+1}]^T$. $n_k$ is a white noise process with variance $\sigma_k^2$. Providing an $\mathbf{x}_k$ and $y_k$ following the above linear system settings, the problem is how to find the weight vector $\mathbf{w}$.

The loss function $J_k$ in the traditional LMS algorithm is regulated as

$$J_k \triangleq e_k^2 / 2 \tag{2}$$

where $e_k$ represents the instantaneous error between the desired response and output, i.e., $e_k \triangleq y_k - \mathbf{w}_k^T \mathbf{x}_k$, where $\mathbf{w}_k \triangleq [w_{k,1}, w_{k,2}, ..., w_{k,N}]^T$ denotes the weight vector of the filter to be estimated at time $k$. With the gradient descent method, the update of classical LMS is then derived as follows

$$\mathbf{w}_{k+1} = \mathbf{w}_k - \mu \frac{\partial J_k}{\partial \mathbf{w}_k} = \mathbf{w}_k + \mu e_k \mathbf{x}_k \tag{3}$$

where the step size $\mu$ satisfies $0 < \mu < \lambda_{\max}^{-1}$ and $\lambda_{\max}$ denotes the maximum eigenvalue of the above $\mathbf{R}$. The gradient descent method applied here guarantees the convergence to the optimum under the aforesaid condition on $\mu$, on account of the convexity of the loss function.

Considering a very common case that the weight vector $\mathbf{w}$ is sparse with the majority of its elements exactly or nearly zeros, many sparsity-aware variants of the classical LMS algorithm have been proposed to improve the performance by employing the prior sparse information. For instance, the $L_0$-LMS, $L_1$-LMS and $L_p$-LMS introduce the $L_0$, $L_1$ and $L_p$ norms of weight vector $\mathbf{w}$, respectively, into the loss function of classical LMS algorithm. For the $L_p$-LMS, the corresponding loss function is

$$J_{k,p} \triangleq e_k^2 / 2 + \gamma_p \|\mathbf{w}_k\|_p \tag{4}$$

where $\|\mathbf{w}_k\|_p \triangleq \left(\sum_{i=1}^N |w_{k,i}|^p\right)^{1/p}$ defines the $L_p$ norm ($0 < p < 1$), with a constant factor $\gamma_p$ weighing the penalty term and determined by the trade-off between the misadjustment and convergence rate. Accordingly, the update is then revised as

$$\mathbf{w}_{k+1} = \mathbf{w}_k + \mu e_k \mathbf{x}_k - \rho_p \frac{\|\mathbf{w}_k\|_p^{1-p} \operatorname{sgn}(\mathbf{w}_k)}{\varepsilon_p + |\mathbf{w}_k|^{1-p}} \tag{5}$$

where $\rho_p = \mu \gamma_p$, $\varepsilon_p$ is a constant imposed to bound the corresponding term in the situation when any component of $\mathbf{w}_k$ approaches zero. The sign function $\operatorname{sgn}(x)$ is defined conventionally, i.e., 1 for $x > 0$, 0 for $x = 0$ and -1 for $x < 0$, and $\operatorname{sgn}(\mathbf{w}_k)$ applies to each element of $\mathbf{w}_k$, respectively. Note that the convergence and consistency analysis of the $L_p$-LMS still remains problematic owing to the non-convexity of its loss function [5].

*Variable $L_p$-LMS Algorithms:* As demonstrated by its proposer Taheri [3], The $L_p$-LMS algorithm can achieve higher performance than the classical LMS, better than or at least comparable to the $L_1$-LMS, $L_0$-LMS and $L_{pl}$-LMS algorithms. However, it suffers the difficulty of adapting to the sparsity of the systems, due to the lack of related adjustable factors. To address this problem, we have developed a new $L_p$-LMS algorithm with a variable $p$, termed as the $L_{vp(GSE)}$-LMS [6], which is achieved by iteratively adjusting $p$ along the opposite direction of the gradient of the instantaneous square error with respect to $p$, leading to an optimal value of $p$ for the $L_p$-LMS and thus a better performance. Since the computational complexity of this approach might be a little high and inspired by the recent research of $L_p$ recovery with affine scaling transformations in compressed sensing [7], we will derive some alternative methods to automatically adjust the value of $p$ of the $L_p$-LMS in the following.

Intuitively, the idea of linearly varying the value of $p$ comes to our sight firstly and simply, i.e.,

$$p_{k+1} = \max\{p_k - s, 0\} \tag{6}$$

where $p_0$ is set to be 1 and the step size $s$ is a very small positive constant, i.e., this method initializes $p$ with the value 1 and then decreases it with a fixed step size $s$. If $p$ reaches 0 before the steady state of convergence, the rest iterations take a fixed $p = 0$ (but we often use a small value like 0.01 instead of zero in simulations).

However, the aforementioned linear variation of $p$ is independent of the iterative algorithm, leading to an improved version that associates the variable $p$ with the convergence preliminarily. Let the deviation $D_k \triangleq \|\mathbf{w}_k - \mathbf{w}_{k-1}\|_2$ representing the $L_2$ norm of the difference between the estimated weight at current iteration $\mathbf{w}_k$ and its previous one $\mathbf{w}_{k-1}$, then we define the relative deviation $RD_k \triangleq D_k / \|\mathbf{w}_k\|_2$ and root RD $rRD_k \triangleq \sqrt{RD_k}$. The improved $p$-variation might be

$$p_{k+1} = \min\{rRD_k, 1\} \tag{7}$$

where the value of $rRD_k$ diminishes and approaches zero at the end of the convergence. If the deviation is lower than a user-defined threshold $\varepsilon$, we will stop the iteration of $p$.

However, the second method doesn't work well since the final value of $p$ is close to 0 which is empirically not the optimal one. It is well known that the LMS algorithm searches the weight vector along the opposite of the gradient of loss function and moves on iteratively with a small step size. Following the similar idea of gradient method, we might vary $p$ as follows

$$p_{k+1} = p_k + \delta \nabla rRD_k / rRD_k \tag{8}$$

where the difference $\nabla rRD_k \triangleq rRD_k - rRD_{k-1}$ denotes the gradient of $rRD_k$ weighted by a positive constant $\delta$. This method, termed as $L_{vpGrRD}$-

LMS, adjusts *p* based on the change in $rRD_k$ and is expected to work effectively with a feedback mechanism.

Additionally, we notice that the above algorithm converges fast as well as keeps stable if the parameter *δ* varies from a larger initial value to a smaller stable value during the iterations. Hence, a simple scheme of a variable *δ* might be given by

$$\delta_{k+1} = \delta_k - u \qquad (9)$$

where the step size *u* is a small positive constant. Moreover, this method might be employ in the $L_{pt}$-LMS as well to improve its performance.

*Simulation results:* In order to verify the performance of the proposed $L_{vpGrRD}$-LMS, numerical simulations are implemented in this section in the light of the convergence rate and steady-state mean square deviation (MSD, $\mathrm{MSD}_k = \mathrm{E}\left[\|\mathbf{w}-\mathbf{w}_k\|_2^2\right]$). Set in the application of sparse system identification, its performances are compared with those of the classical LMS and $L_p$-LMS algorithms in different sparsity levels.

The sparse system to be estimated has 16 taps, whose 1, 4, 8 or 16 tap(s) are established to be non zeros, setting the sparsity ratio (SR) as 1/16, 4/16, 8/16 and 16/16, respectively. The positions of the nonzero taps are chosen randomly with their values setting to be 1 and -1 randomly as well. The noise signal and input signal employed are both assumed to be white Gaussian processes of a fixed length 500 with variances 1 and 0.01 (or 0.001), respectively, leaving the signal-to-noise ratio (SNR) 20 dB or 30 dB. For the sake of concisely presentation, SR = 1/16 is set for the 1~500th iteration, SR = 4/16 for the following 500 ones, and lastly SR = 8/16 for the left 500 iterations. Listed in Table 1 are all the left parameters that are discreetly and fairly selected. All the simulation results are averaged for 200 times. Note that the value of *p* varies from the 10th to the 200th iterations with different parameter *δ*'s under different SRs in the experiment.

Table 1. Residual parameters of different approaches in this simulations.

| Approaches | *μ* | *ε* | *ρ* | *p* | *δ* |
|---|---|---|---|---|---|
| LMS | | / | / | / | / |
| $L_p$-LMS | 0.05 | 0.05 | $5\times10^{-5}$ | 0.5 | / |
| $L_{vpGrRD}$-LMS | | | | $p_0$=1 | 0.02* |
| * Different *δ* = 0.02, 0.02, 0.01 and 0.005 are intended for SR = 1/16, 4/16, 8/16 and 16/16, separately. | | | | | |

Fig. 1 presents three MSD curves of the tested algorithms with horizontal axis the number of iterations at different sparsity levels, viz. SR = 1/16, 4/16 and 8/16. As can be seen from Fig.1, the $L_{vpGrRD}$-LMS generally yields significantly lower steady-state error than both the classical LMS and $L_p$-LMS do with different sparsity ratios, due to the adjustment of the value of *p*. Specifically, the stable MSD of $L_{vpGrRD}$-LMS is much higher than the $L_p$-LMS in the very sparse case, i.e., SR = 1/16, and they both outperform the classical LMS with the help of sparse penalty, i.e., the $L_p$ norm constraint; When SR increases, the proposed $L_{vpGrRD}$-LMS algorithm still achieves lower stable error than the $L_p$-LMS and classical LMS, except in case that the system is totally non-sparse (SR = 16/16, see Fig. 2) when the $L_{vpGrRD}$-LMS performs very close to the classical LMS, way better than the $L_p$-LMS that deteriorates in non-sparse cases as expected. Correspondingly, the iteration of the varying *p* is presented in Fig. 3, showing that *p* iteratively converges from the initial value *p* = 1 for all different SRs except the last non-sparse one.

*Conclusion:* In order to further improve the performance of the system identification with the popular $L_p$-LMS algorithm, the $L_{vpGrRD}$-LMS approach was proposed as a supplement of the former where the variable parameter *p* was iteratively adapted to the gradient of root relative deviation of the estimate weight vector. Numerical simulations showed that this proposed algorithm achieves lower steady-state error and equally high rate of convergence than the traditional $L_p$-LMS and classical LMS algorithm.

*Acknowledgments:* This work was supported by the NSFC (grants 61571213 and 61201344).

Yong Feng and Fei Chen (*Department of Electrical and Electronic Engineering, South University of Science and Technology of China, Shenzhen 518055, China*)
✉ E-mail: fengy@sustc.edu.cn

Jiasong Wu (*School of Computer Science and Engineering, Southeast University, Nanjing 210096, China*)

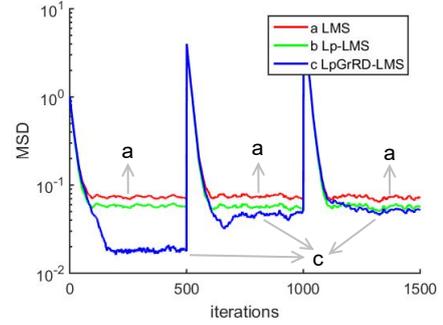

Fig. 1 *MSD curves of the three tested approaches with SNR = 20 dB and SR = 1/16, 4/16 or 8/16, respectively.*

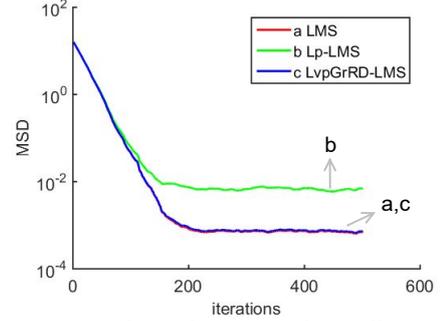

Fig. 2 *The iterations of p in the setting of a totally non-sparse system (SR = 16/16, SNR = 30dB).*

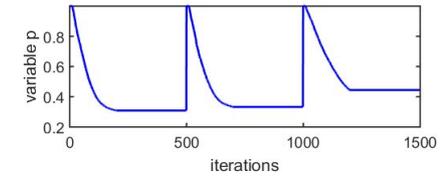

Fig. 3 *The iterations of p under different SRs (corresponding to Fig.1).*